\documentclass[13 pt]{article}
\usepackage{amssymb}

\newcommand{\beq}{\begin{equation}}
\newcommand{\eeq}{\end{equation}}
\newcommand{\bea}{\begin{eqnarray*}}
\newcommand{\eea}{\end{eqnarray*}}
\newcommand{\beqa}{\begin{eqnarray}}
\newcommand{\eeqa}{\end{eqnarray}}

\begin{document}

\newfont{\elevenmib}{cmmib10 scaled\magstep1}%

\newcommand{\Title}[1]{{\baselineskip=26pt \begin{center}
            \Large   \bf #1 \\ \ \\ \end{center}}}
\hspace*{2.13cm}%
\hspace*{1cm}%
\newcommand{\Author}{\begin{center}\large
           Pascal Baseilhac\footnote{
baseilha@phys.univ-tours.fr} and Kozo Koizumi\footnote{kozo.koizumi@lmpt.univ-tours.fr}
\end{center}}
\newcommand{\Address}{{\baselineskip=18pt \begin{center}
           \it Laboratoire de Math\'ematiques et Physique Th\'eorique CNRS/UMR 6083,\\
Universit\'e de Tours, Parc de Grandmont, 37200 Tours, France
      \end{center}}}
\baselineskip=13pt

\bigskip
\vspace{-1cm}

\Title{A deformed analogue of Onsager's symmetry\\ in the XXZ open spin chain}\Author

\vspace{- 0.1mm}
 \Address

\vskip 0.6cm

\centerline{\bf Abstract}\vspace{0.3mm}  \vspace{1mm}
The XXZ open spin chain with general integrable boundary conditions is shown to possess a $q-$deformed analogue of the Onsager's algebra as fundamental non-abelian symmetry which ensures the integrability of the model. This symmetry implies the existence of a finite set of independent mutually commuting nonlocal operators which form an abelian subalgebra. The transfer matrix and local conserved quantities, for instance the Hamiltonian, are expressed in terms of these nonlocal operators. It follows that Onsager's original approach of the planar Ising model can be extended to the XXZ open spin chain. 
\vspace{0.1cm} 

\vskip 0.8cm

\vskip -0.6cm


\section{Introduction}
Quantum integrable models have always attracted attention, starting from Bethe's work \cite{Bethe} on Heisenberg isotropic spin chain, and Onsager's solution of the planar Ising model \cite{Ons}. They possess several important domain of applications which range from (super)string theory to condensed matter systems, and allow to obtain exact and nonperturbative expressions for physical quantities (scattering processes, correlation functions,...). Furthermore, they are deeply related with nice and rich mathematical structures (like the theory of quantum Lie algebras, $q-$orthogonal polynomials, partial differential and $q-$difference equations,...).
In recent years, much interest has been focused on quantum integrable models with boundaries \cite{Cher,Sklya} for which several exact results have already been obtained although a large number of questions remain. For instance, the underlying symmetry of these models, exact spectrum and correlation functions are yet to be found. \vspace{1mm}

Among these models, one finds for instance the XXZ anisotropic open spin chain with general integrable boundary conditions \cite{Sklya,DeVeg} which has received a lot of attention in the past years \cite{Nepo,Cao,Rittenberg}. Despite of - nevertheless important - partial results (some symmetries for a special choice of boundary parameters, spectrum and Bethe ansatz equations for special relations between left and right boundary parameters, a connection with the Ising model at special values of the anisotropy parameter and special diagonal boundary conditions \cite{Bel}), its underlying symmetry for arbitrary boundary parameters clearly needs to be clarified. Indeed, for special diagonal boundary conditions at both ends of the spin chain the model is known to enjoy $U_{q^{1/2}}(sl_2)$ symmetry \cite{Pasquier}. Also, for special boundary conditions (left identical non-diagonal and right identical diagonal) it was observed in \cite{Doikou} that some nonlocal operators\,\footnote{Nonlocal in the sense that they act simulateneously on all sites of the chain. Note that the structure of these operators is consistent with the one derived in the context of quantum field theory for the boundary sine-Gordon model \cite{Nepo1,MacDel}.} act non-trivially on the double-row monodromy matrix (also called Sklyanin operator) and Hamiltonian of the open spin chain.  Despite of these facts, for non-diagonal boundary parameters neither the {\it non-abelian} algebra on which these nonlocal operators close (the boundary analogue of $U_{q^{1/2}}(\widehat{sl_2})$), nor its abelian subalgebra (generating all integrals of motion and implying the integrability of the system) have been explicitly identified. This might be even more surprising, as the term ``boundary quantum group algebra'' corresponding to such algebraic structure is used repeatedely but with no reference to proper defining relations independent of the spectral parameter $u$, derived from the reflection equation\,\footnote{By analogy, the $U_q({\widehat{sl_2}})$ algebra is associated with a set of fundamental defining relations independent of $u$, although derived from the ``$RTT$'' relations (Yang-Baxter algebra).}. However, a breakthrough recently emerged \cite{qDG,TriDiag} from the study of the algebraic structure encoded in the reflection equation for certain (left purely non-diagonal and right diagonal) boundary conditions and $U_{q^{1/2}}(\widehat{sl_2})$ trigonometric $R-$matrix. Indeed, in \cite{qDG,TriDiag} it was discovered that the  boundary analogue of $U_{q^{1/2}}(\widehat{sl_2})$ is associated with the tridiagonal algebra, first introduced by Terwilliger \cite{Ter}. This algebra is defined through a pair of {\it tridiagonal} relations; these are extensions of   the $q-$Serre relations and can be viewed as ``$q-$deformed'' Dolan-Grady relations due to their close analogy with the relations discovered in \cite{DG}. In addition, later on it was argued \cite{qOns} that this algebra  is related with a $q-$deformed analogue of the Onsager's algebra\,\footnote{
For $q=1$ and special value of $\rho$, the algebra introduced in \cite{qOns} can be related with the Onsager's algebra which played a crucial role in the original solution of the planar Ising model (see \cite{Ons} for details).
} constructed in \cite{qOns}. As briefly mentionned in \cite{TriDiag,qOns} this remarkable tridiagonal algebraic structure survives for the most general (non-diagonal) solutions of the reflection equation.\vspace{1mm}

In this paper, we show that the homogeneous XXZ open spin-$\frac{1}{2}$ chain with the most general integrable boundary conditions can be built from a finite set of nonlocal operators which generate a $q-$deformed analogue of the Onsager's algebra and ensure the integrability of the system. The symmetry of the model is identified. Then, our results open the possibility to apply Onsager's original approach \cite{Ons} to integrable open spin chains.\vspace{1mm}

The paper is organized as follows. In the next Section, it is shown that the entries of the double-row monodromy matrix (Sklyanin operator) can be written in terms of a finite set of nonlocal operators acting on finite dimensional tensor product representations and generating a $q-$deformed analogue of the Onsager's algebra. Then,
we study in details the abelian symmetry of the model: the Hamiltonian is shown to commute with a finite set of nonlocal operators ${\cal I}_{2k+1}^{(N)}$ which generate an abelian subalgebra of the $q-$deformed Onsager's algebra. 
In the last Section, based on the knowledge of the symmetry we argue that the Onsager's approach can be applied to the XXZ open spin chain. Indeed, we explain that the spectral problem for the transfer matrix, and in particular the Hamiltonian, reduces to the one for the nonlocal operators ${\cal I}_{2k+1}^{(N)}$, $k\in 0,...,N-1$. 

\section{The XXZ open spin chain with integrable boundary conditions}
The XXZ open spin chain with general integrable boundary conditions has recently attracted much attention. This started from the work of Sklyanin \cite{Sklya}, who proposed a systematic procedure to find boundary conditions for open spin chain compatible with integrability. Since this work, the objective remains the same: derive exact results such as the energy spectrum of the open spin chain for arbitrary integrable boundary conditions and find exact correlation functions, extending the known results for the infinite volume limit (see for instance \cite{Jimbo} and references therein). In particular, in the infinite volume limit the identification of $U_{q^{1/2}}({\widehat{sl_2}})$ as the underlying symmetry of the spin chain played a crucial role in deriving such examples of exact results.
Similarly, identifying the underlying symmetry of the XXZ open spin$-\frac{1}{2}$ chain for general integrable boundary conditions 
is a very important problem, which is completely solved in this Section. In the following, we consider the Hamiltonian
\beqa
H_{XXZ}&=&\sum_{k=1}^{N-1}\Big(\sigma_1^{k}\sigma_1^{k+1}+\sigma_2^{k}\sigma_2^{k+1} + \Delta\sigma_3^{k}\sigma_3^{k+1}\Big) \nonumber\\
&&+\ \frac{(q^{1/2}-q^{-1/2})}{(\epsilon^{(0)}_+ + \epsilon^{(0)}_-)}
\Big(  \frac{(\epsilon^{(0)}_+ - \epsilon^{(0)}_-)}{2}\sigma^1_3 + \frac{2}{(q^{1/2}-q^{-1/2})}\big(k_+\sigma^1_+ + k_-\sigma^1_-\big)       \Big)\nonumber\\
 &&+\ \frac{(q^{1/2}-q^{-1/2})}{(\kappa + \kappa^*)}
\Big(  \frac{(\kappa - \kappa^*)}{2}\sigma^N_3 + 2(q^{1/2}+q^{-1/2})\big(\kappa_+\sigma^N_+ + \kappa_-\sigma^N_-\big)       \Big)\label{H}\ ,
\eeqa
where $\sigma_{1,2,3}$ and $\sigma_\pm=(\sigma_1\pm i\sigma_2)/2$ are usual Pauli matrices. Here, $\Delta=(q^{1/2}+q^{-1/2})/2$\ \ denotes the anisotropy parameter. Boundary parameters associated with the most general integrable boundary conditions are such that
\beqa
&&\epsilon^{(0)}_{+}=(\epsilon^{(0)}_{-})^{\dagger}=(c_{00}+ic_{01})/2\ ,\qquad k_+=(k_{-})^{\dagger}=-(q^{1/2}-q^{-1/2})e^{i\theta}/2 \qquad\ \  \mbox{(left)}\ ,\nonumber\\ 
&&\kappa=(\kappa^*)^{\dagger}=(-{\tilde{c}}_{00}+i{\tilde{c}}_{01})/2\ ,\qquad
\kappa_+=(\kappa_{-})^{\dagger}=-e^{i{\tilde{\theta}}}/(2(q^{1/2}+q^{-1/2}))\qquad \mbox{(right)}\ .\label{param}
\eeqa
In order exhibit independent parameters, the r.h.s parametrization can be replaced in (\ref{H}) in which case
one immediatly identifies the Hamiltonian as defined in \cite{Cao}. In total, one has {\it six} boundary parameters $c_{00},c_{01},{\tilde{c}}_{00},{\tilde{c}}_{01},\theta, {\tilde{\theta}}$. Considering a gauge transformation, note that one parameter might be removed in which case only five arbitrary parameters remain. For symmetry reasons, we however keep the boundary parametrization as defined above. Note that restricting the boundary parameters to special values, one obtains the cases considered in \cite{Sklya,Rittenberg,Doikou}. \vspace{1mm}

A characteristic feature of quantum integrable systems is the existence of an infinite number of mutually commuting integrals of motion. For quantum integrable systems on the lattice with non-periodic boundary conditions, Sklyanin \cite{Sklya} proposed a rather general method to derive in a systematic way this family of independent mutually commuting quantities. Similarly to the periodic case, a transfer matrix which provides the generating functional for all integrals of motion can be constructed. For the homogeneous XXZ open spin$-\frac{1}{2}$ chain with $N$ sites, this transfer matrix takes the form \cite{Sklya}
\beqa
t_{XXZ}(u)=\frac{(-1)^N}{(u^2+u^{-2}-q-q^{-1})^N}\  tr_0\Big[K_+(u)(id \otimes_{k=1}^{N}\pi^{(\frac{1}{2})})[K^{(N)}(u)]|_{v_1,...,v_N=1}\Big]\ ,\label{tXXZ}
\eeqa
where
$tr_{0}$ denotes the trace over the two-dimensional
auxiliary space, the Sklyanin operator
\beqa K^{(N)}(u)\equiv L_{\verb"N"}(uv_N)\cdot\cdot\cdot
L_{\verb"1"}(uv_1)K_-(u)L_{\verb"1"}(uv_1^{-1})\cdot\cdot\cdot
L_{\verb"N"}(uv_N^{-1})\ \label{Kdressed}\eeqa
defines\,\footnote{Note that $K^{(N)}(u)$ acts on a tensor product representation of ``quantum'' spaces. Standard matrix multiplication over the auxiliary space is used to multiply the $L-$operators.} a family of solution to the reflection equation
\beqa R(u/v)\ (K^{(N)}(u)\otimes I\!\!I)\ R(uv)\ (I\!\!I \otimes K^{(N)}(v))\
= \ (I\!\!I \otimes K^{(N)}(v))\ R(uv)\ (K^{(N)}(u)\otimes I\!\!I)\ R(u/v)\ 
\label{RE} \eeqa
for arbitrary parameters $v_1,...,v_N\in {\mathbb C}$ and $\pi^{(\frac{1}{2})}$ induces a spin$-\frac{1}{2}$ representation for the quantum space located at each site of the spin chain. As shown in \cite{Sklya}, the integrability relies on the fact that given a $R-$matrix the Lax operator $L(u)$ satisfies the Yang-Baxter algebra, $K_-(u)$ in (\ref{Kdressed}) is {\it itself} a solution of (\ref{RE}) and $K_{+}(u)$ solves the ``dual'' reflection equation\,\footnote{The ``dual'' reflection equation 
follows from (\ref{RE}) by changing $u\rightarrow
q^{-1/2}u^{-1}$, $v\rightarrow q^{-1/2}v^{-1}$ and $K(u)$ in its transpose.}. In the context of integrable open spin chains, the element $K_-(u)$ and $K_+(u)$, respectively, characterize the class of left (resp. right) integrable boundary conditions. For the XXZ open spin chain with transfer matrix (\ref{tXXZ}), given the trigonometric $R-$matrix associated with $U_{q^{1/2}}(\widehat{sl_2})$ 
\beqa {R}(u) =\left(
\begin{array}{cccc}
 uq^{{1/2}}- u^{-1}q^{-{1/2}}    & 0 & 0 & 0 \\
 0 & u- u^{-1} & q^{{1/2}}- q^{-{1/2}} & 0\\
0 & q^{1/2}-q^{-1/2} &  u- u^{-1} & 0  \\
0 & 0 & 0 & uq^{{1/2}}- u^{-1}q^{-{1/2}} 
\end{array} \right) \ ,\label{R}
\eeqa
the most general elements $K_\pm(u)$ with $c-$number entries take the form
\beqa K_-(u) &=&\left(
\begin{array}{cc} 
 u \epsilon^{(0)}_{+} +  u^{-1}\epsilon^{(0)}_{-}     &   k_+(u^2-u^{-2})/(q^{1/2}-q^{-1/2})\\
k_-(u^2-u^{-2})/(q^{1/2}-q^{-1/2})   &  u \epsilon^{(0)}_{-} +  u^{-1}\epsilon^{(0)}_{+}  \\
\end{array} \right) \ ,\label{K-}\\
K_+(u) &=&\left(
\begin{array}{cc} 
 q^{1/2}u\kappa +  q^{-1/2}u^{-1}\kappa^*     &   \kappa_+(qu^2-q^{-1}u^{-2})(q^{1/2}+q^{-1/2})\\
\kappa_-(qu^2-q^{-1}u^{-2})(q^{1/2}+q^{-1/2})  &  q^{1/2}u \kappa^* +  q^{-1/2}u^{-1}\kappa  \\
\end{array} \right) \ \label{K+}
\eeqa
with (\ref{param}), respectively. For further convenience, note that we have chosen a different normalization in the off-diagonal boundary terms between (\ref{K+}) and (\ref{K-}).\vspace{1mm}

With the definitions above, differentiating the transfer matrix (\ref{tXXZ})  $n-$times with respect to $\lambda=\ln(u)$ at $\lambda=0$ (i.e. $u=1$),  mutually commuting {\it local} conserved quantities can be derived. Recall that 
\beqa
R_{0k}(u=1)=(q^{1/2}-q^{-1/2}){\cal P}_{0k}\ ,\nonumber
\eeqa
where ${\cal P}_{kl}$ denotes the permutation operator with properties
\beqa
{\cal P}_{kl}={\cal P}_{lk}\ ,\quad {\cal P}^2_{kl}=1\ , \quad {\cal P}_{kl}F_{lm}=F_{km}{\cal P}_{kl} \quad m\neq k,l\ \nonumber
\eeqa
for an arbitrary operator $F_{kl}$ acting on sites $k$ and $l$. Then, for $n=1$ one derives in a straightforward manner
\beqa
\frac{d}{du} ln(t_{XXZ}(u))|_{u=1}= \Big(\frac{(q^{1/2}-q^{-1/2})}{(q^{1/2}+q^{-1/2})} + \frac{2N}{(q^{1/2}-q^{-1/2})}\Delta \Big)I\!\!I^{(N)} + \frac{2}{(q^{1/2}-q^{-1/2})}H_{XXZ}\label{expH}\ ,
\eeqa
where the Hamiltonian given by (\ref{H}) is one of the local conserved quantities. More generally, higher mutually commuting local conserved quantities, say $H_n$ with $H_1\equiv H_{XXZ}$ can be derived similarly.

\subsection{Preliminaries: structure of the Sklyanin operator}
The Sklyanin operator, as defined in (\ref{Kdressed}) is one of the fundamental element in the transfer matrix of the XXZ open spin chain (\ref{tXXZ}) associated with the Hamiltonian (\ref{H}).  For further analysis, let us write it in the form
\beqa K^{(N)}(u) =\left(
\begin{array}{cc} 
    {\cal A}^{(N)}(u)  &  {\cal B}^{(N)}(u)  \\
  {\cal C}^{(N)}(u)  & {\cal D}^{(N)}(u)    \\
\end{array} \right) \ .\label{SN} \eeqa
Remarkably \cite{TriDiag}, it is possible to {\it systematically} disentangle for arbitrary integer $N$ the algebraic part   (acting on tensor product representations of $N$ quantum spaces) from the spectral parameter (i.e. dependent on ``$u$'') part. Indeed, using the procedure presented in \cite{TriDiag}, for (\ref{K-}), (\ref{K+}) and arbitrary $N$ after some rather involved calculations (see \cite{TriDiag,qOns} for details) the Sklyanin operator (\ref{SN}) is such that 
\beqa
{\cal A}^{(N)}(u) &=&  u \epsilon^{(N)}_{+} +  u^{-1}\epsilon^{(N)}_{-} + (u^2-u^{-2})\Big(uq^{1/2}\sum_{k=0}^{N-1} P_{-k}^{(N)}(u)
{\textsf W}_{-k}^{(N)}-u^{-1}q^{-1/2}\sum_{k=0}^{N-1} P_{-k}^{(N)}(u) {\textsf W}_{k+1}^{(N)}\Big)\ ,\nonumber\\
{\cal D}^{(N)}(u) &=&  u \epsilon^{(N)}_{-} +  u^{-1}\epsilon^{(N)}_{+} + (u^2-u^{-2})\Big(uq^{1/2}\sum_{k=0}^{N-1} P_{-k}^{(N)}(u)
{\textsf W}_{k+1}^{(N)}-u^{-1}q^{-1/2}\sum_{k=0}^{N-1} P_{-k}^{(N)}(u) {\textsf W}_{-k}^{(N)}\Big)\ ,\nonumber\\ 
{\cal B}^{(N)}(u) &=& \frac{(u^2-u^{-2})}{k_-}\Big(\frac{k_+k_-(q^{1/2}u^2+q^{-1/2}u^{-2})}{(q^{1/2}-q^{-1/2})}P_0^{(N)}(u)
+\frac{1}{q^{1/2}+q^{-1/2}}\sum_{k=0}^{N-1}P_{-k}^{(N)}(u){\textsf G}_{k+1}^{(N)}+{\omega}_0^{(N)}\ \Big)\ ,\nonumber\\
{\cal C}^{(N)}(u)&=&\frac{(u^2-u^{-2})}{k_+}\Big(\frac{k_+k_-(q^{1/2}u^2+q^{-1/2}u^{-2})}{(q^{1/2}-q^{-1/2})}P_0^{(N)}(u)
+\frac{1}{q^{1/2}+q^{-1/2}}\sum_{k=0}^{N-1}P_{-k}^{(N)}(u){\tilde
{\textsf G}}_{k+1}^{(N)}+{\omega}_0^{(N)}\Big)\ . \label{KN}
\eeqa
Compared to (\ref{Kdressed}), although apparently complicated this expression for $K^{(N)}(u)$ possesses rather nice features. It contains non-algebraic ($c-$number valued) terms such as the Laurent polynomials $P_{-k}^{(N)}(u)$,
$\epsilon^{(N)}_{\pm}$, ${\omega}_0^{(N)}$ separated from ``$u$''- independent operators   ${\textsf W}_{-k},{\textsf W}_{k+1},{\textsf G}_{k+1},{\tilde
{\textsf G}}_{k+1}$. Namely, one finds\vspace{1mm}

$\bullet$ {\bf Non-algebraic part:} The ``initial'' solution $K_-(u)$ of the reflection equation (\ref{RE}) being slightly different from the choice in \cite{qOns} (here, diagonal boundary parameters $\epsilon^{(0)}_\pm$ are considered), the final expressions for the Laurent polynomials become
\beqa
P_{-k}^{(N)}(u)=-\frac{1}{(q^{1/2}+q^{-1/2})}
\sum_{n=k}^{N-1}\Big(\frac{q^{1/2}u^2+q^{-1/2}u^{-2}}{q^{1/2}+q^{-1/2}}\Big)^{n-k}C_{-n}^{(N)}, \label{PN}
\eeqa
\beqa
C_{-n}^{(N)}=(-1)^{N-n} (q^{1/2}+q^{-1/2})^{n+1}\sum_{k_1<...<k_{N-n-1}=1}^{N}\alpha_{k_1}\cdot \cdot \cdot\alpha_{k_{N-n-1}}
\ \nonumber
\eeqa
where we have defined
\beqa
\alpha_1&=& \frac{(v_{1}^2+v_{1}^{-2})w_0^{(j_{1})}}{(q^{1/2}+q^{-1/2})}+ \frac{\epsilon^{(0)}_{+}\epsilon^{(0)}_{-}(q^{1/2}-q^{-1/2})^2}{k_+k_-(q^{1/2}+q^{-1/2})}\ ,\nonumber\\ \alpha_{k}&=&\frac{(v_{k}^2+v_{k}^{-2})w_0^{(j_{k})}}{(q^{1/2}+q^{-1/2})}\quad \mbox{for}\quad k=2,...,N\ .\nonumber
\eeqa
The remaining constant terms in (\ref{KN}) are
\beqa
\epsilon^{(N)}_{\pm}=w_0^{(j_{N})}\epsilon^{(N-1)}_{\mp} - (v_{N}^2+v_{N}^{-2})\epsilon^{(N-1)}_{\pm}\qquad \mbox{and} \qquad \omega_0^{(N)}=(-1)^{N}\frac{k_+k_-}{q^{1/2}-q^{-1/2}}\prod_{k=1}^{N}\alpha_k\ 
\eeqa
for arbitrary values of $N$. In all expressions above,
\beqa
w_0^{(j_k)}=q^{j_k+1/2}+q^{-j_k-1/2}\nonumber
\eeqa
denotes the Casimir operator eigenvalue associated with the spin$-j_k$ representation of $U_{q^{1/2}}(\widehat{sl_2})$ for $k=1,...,N$. Note that setting $\epsilon^{(0)}_\pm=0$, $j_k=j$, $k_+=1/c_0$, $k_-=1$ and $v_k=v$ in above expressions, one recovers the results of \cite{qOns}.\vspace{1mm}

$\bullet$ {\bf Algebraic part:}  For general values\footnote{To understand the general case $N$, it is instructive to look at the special cases $N=1$ and $N=2$ first \cite{TriDiag,qOns}. 
For $N=1$ in (\ref{Kdressed}), it is easy to show that ${\textsf W}_{0},{\textsf W}_{1}$ are simply given by linear combinations of $U_{q^{1/2}}(\widehat{sl_2})$ generators $S_\pm q^{\pm s_3/2},q^{\pm s_3}$ with coefficients depending on the parameter $v_1$, whereas ${\textsf G}_{1},{\tilde
{\textsf G}}_{1}$ are linear combinations of $S^2_\pm,q^{\pm s_3}$ and $S_\pm q^{\pm s_3/2}$ (for non-vanishing values of $\epsilon^{(0)}_\pm$). 
For $N=2$, the elements ${\textsf W}_{0},{\textsf W}_{-1},{\textsf W}_{1},{\textsf W}_{2}$ and ${\textsf G}_{1},{\textsf G}_{2},{\tilde
{\textsf G}}_{1},{\tilde {\textsf G}}_{2}$ are linear combination of tensor product of $U_{q^{1/2}}(\widehat{sl_2})$ generators, with coefficients depending on $v_1,v_2$.}
of $N$, the elements ${\textsf W}_{-k},{\textsf W}_{k+1},{\textsf G}_{k+1},{\tilde
{\textsf G}}_{k+1}$ ($4N$ in total) act on $N-$tensor product representations of $U_{q^{1/2}}(\widehat{sl_2})$, and depend solely on the $N-$parameters $v_k$ and spin$-j_k$ for $k=1,...,N$. Following an analysis similar to \cite{qOns} for (\ref{K-}), (\ref{K+}) one eventually gets the family of nonlocal operators which act as:
\beqa
{\textsf W}_{-k}^{(N)}&=&\frac{(w_0^{(j_{N})}-(q^{1/2}+q^{-1/2})q^{s_3})}{(q^{1/2}+q^{-1/2})}\otimes
{\textsf W}_{k}^{(N-1)}
-\frac{(v_{N}^2+v_{N}^{-2})}{(q^{1/2}+q^{-1/2})}I\!\!I\otimes {\textsf W}_{-k+1}^{(N-1)} +\ \frac{(v_N^2+v_N^{-2})w_0^{(j_N)}}{(q^{1/2}+q^{-1/2})^2}{\textsf W}_{-k+1}^{(N)}
\nonumber\\
&&\ \ \ + \ \frac{(q^{1/2}-q^{-1/2})}{k_+k_-(q^{1/2}+q^{-1/2})^2}
\left(k_+v_Nq^{1/4}S_+q^{s_3/2}\otimes
{\textsf G}_{k}^{(N-1)}+k_-v_N^{-1}q^{-1/4}S_-q^{s_3/2}\otimes {\tilde {\textsf G}}_{k}^{(N-1)}\right)\nonumber\\
&&\ \ \ +\ q^{s_3}\otimes {\textsf W}_{-k}^{(N-1)}
\ ,\nonumber\\
{\textsf W}_{k+1}^{(N)}&=&\frac{(w_0^{(j_{N})}-(q^{1/2}+q^{-1/2})q^{-s_3})}{(q^{1/2}+q^{-1/2})}\otimes
{\textsf W}_{-k+1}^{(N-1)}
-\frac{(v_{N}^2+v_{N}^{-2})}{(q^{1/2}+q^{-1/2})}I\!\!I\otimes {\textsf W}_{k}^{(N-1)} +\ \frac{(v_N^2+v_N^{-2})w_0^{(j_N)}}{(q^{1/2}+q^{-1/2})^2}{\textsf W}_{k}^{(N)}
\nonumber\\
&&\ \ \ +\ \frac{(q^{1/2}-q^{-1/2})}{k_+k_-(q^{1/2}+q^{-1/2})^2}
\left(k_+v^{-1}_Nq^{-1/4}S_+q^{-s_3/2}\otimes
{\textsf G}_{k}^{(N-1)}+k_-v_Nq^{1/4}S_-q^{-s_3/2}\otimes {\tilde {\textsf G}}_{k}^{(N-1)}\right)\nonumber\\
&&\ \ \ +\ q^{-s_3}\otimes {\textsf W}_{k+1}^{(N-1)}
\ ,\nonumber\\
{\textsf G}_{k+1}^{(N)}&=& 
\frac{k_-(q^{1/2}-q^{-1/2})^2}{k_+(q^{1/2}+q^{-1/2})}
S_-^2\otimes {\tilde {\textsf G}}_{k}^{(N-1)}
-\frac{1}{(q^{1/2}+q^{-1/2})}(v_N^{2}q^{s_3}+v_N^{-2}q^{-s_3})\otimes {\textsf G}_{k}^{(N-1)} 
+I\!\!I \otimes {\textsf G}_{k+1}^{(N-1)}\nonumber\\
&& + (q-q^{-1})\left(
k_-v_Nq^{-1/4}S_-q^{s_3/2}\otimes \big({\textsf W}_{-k}^{(N-1)}-{\textsf W}_{k}^{(N-1)}\big)
+k_-v_N^{-1}q^{1/4}S_-q^{-s_3/2}\otimes \big({\textsf W}_{k+1}^{(N-1)}-{\textsf W}_{-k+1}^{(N-1)}\big)
\right)\nonumber\\
&&+\frac{(v_N^2+v_N^{-2})w_0^{(j_N)}}{(q^{1/2}+q^{-1/2})^2}{\textsf G}_{k}^{(N)}\ ,\nonumber\\
\nonumber\\
{\tilde{\textsf G}}_{k+1}^{(N)}&=& 
\frac{k_+(q^{1/2}-q^{-1/2})^2}{k_-(q^{1/2}+q^{-1/2})}
S_+^2\otimes {{\textsf G}}_{k}^{(N-1)}
-\frac{1}{(q^{1/2}+q^{-1/2})}(v_N^{2}q^{-s_3}+v_N^{-2}q^{s_3})\otimes {\tilde{\textsf G}}_{k}^{(N-1)} 
+I\!\!I \otimes {\tilde{\textsf G}}_{k+1}^{(N-1)}\nonumber\\
&& + (q-q^{-1})\left(
k_+v^{-1}_Nq^{1/4}S_+q^{s_3/2}\otimes \big({\textsf W}_{-k}^{(N-1)}-{\textsf W}_{k}^{(N-1)}\big)
+k_+v_Nq^{-1/4}S_+q^{-s_3/2}\otimes \big({\textsf W}_{k+1}^{(N-1)}-{\textsf W}_{-k+1}^{(N-1)}\big)
\right)\nonumber\\
&&+\frac{(v_N^2+v_N^{-2})w_0^{(j_N)}}{(q^{1/2}+q^{-1/2})^2}{\tilde{\textsf G}}_{k}^{(N)}\ ,\label{rep}
\eeqa
where, for the special case $k=0$ we identify\,\footnote{Although the notation is ambiguous, the reader must keep in mind that ${{\textsf W}}_{k}^{(N)}|_{k=0}\neq {{\textsf W}}_{-k}^{(N)}|_{k=0}$\ ,${{\textsf W}}_{-k+1}^{(N)}|_{k=0}\neq {{\textsf W}}_{k+1}^{(N)}|_{k=0}$\ for any $N$.}
\beqa
{{\textsf W}}_{k}^{(N)}|_{k=0}\equiv 0\ ,\quad {{\textsf W}}_{-k+1}^{(N)}|_{k=0}\equiv 0\ ,\quad {\textsf G}_{k}^{(N)}|_{k=0}={\tilde{\textsf G}}_{k}^{(N)}|_{k=0}\equiv \frac{k_+k_-(q^{1/2}+q^{-1/2})^2}{(q^{1/2}-q^{-1/2})}I\!\!I^{(N)}\ .\label{not}
\eeqa
Here, the identity operator $I\!\!I^{(N)}=I\!\!I\otimes \cdot \cdot \cdot \otimes I\!\!I$
acting on the $N-$quantum spaces tensor product 
has been introduced. In addition, one has the ``initial'' $c-$number conditions
\beqa
{{\textsf W}}_{0}^{(0)}\equiv \epsilon^{(0)}_+\ ,\quad {{\textsf W}}_{1}^{(0)}\equiv \epsilon^{(0)}_-\qquad  \mbox{and}\qquad
{\textsf G}_{1}^{(0)}={\tilde{\textsf G}}_{1}^{(0)}\equiv \epsilon^{(0)}_+\epsilon^{(0)}_-(q^{1/2}-q^{-1/2})\ .\label{initrep}
\eeqa

For the sake of completness, let us mention some important properties of the elements (\ref{rep}) given above, which can be deduced inserting $K^{(N)}(u)$ in the reflection equation (\ref{RE}). Expanding in the spectral parameter $u$, one obtains a pair of relations linear in terms of the elements  ${\textsf W}_{-k},{\textsf W}_{k+1}$ which ensures the closure of the algebra. For $N=1$, they correspond to the Askey-Wilson relations \cite{qDG,TriDiag}, trilinear in terms of ${\textsf W}_{0},{\textsf W}_{1}$ or, equivalently, linear in terms of ${\textsf W}_{0},{\textsf W}_{1},{\textsf W}_{-1},{\textsf W}_{2}$ in which case ${\textsf W}_{-1},{\textsf W}_{2}$ are defined as nonlinear combinations of ${\textsf W}_{0},{\textsf W}_{1}$. 
For $N=2$, one obtains relations of fifth-order in ${\textsf W}_{0},{\textsf W}_{1}$, or linear in terms of the fundamental elements ${\textsf W}_{-k},{\textsf W}_{k+1}$, $k=0,1,2$  having identified ${\textsf W}_{-2},{\textsf W}_{3}$ with certain nonlinear combinations of  ${\textsf W}_{0},{\textsf W}_{1}$ \cite{qOns}. In general, given $N$ a pair of relations of order $2N+1$ in ${\textsf W}_{0},{\textsf W}_{1}$ exists, which can be written as a pair of linear recurrence relation of length $N+1$ in the fundamental elements ${\textsf W}_{-k},{\textsf W}_{k+1}$, $k=0,1,...,N$ where ${\textsf W}_{-N},{\textsf W}_{N+1}$ are nonlinear combinations of lower elements ${\textsf W}_{-k},{\textsf W}_{k+1}$ with $k=0,1,...,N-1$. In the present work, for the choice (\ref{K-}), (\ref{K+}) straightforward calculations similar to \cite{qOns} yield to\,\footnote{More general relations can be obtained as a consequence of these fundamental ones. For details, we refer the reader to \cite{qOns}, Appendix B. For $q=1$, note that the relations (\ref{c2}), (\ref{c4}) as well as their generalizations \cite{qOns} reduce to the finite recurrence relations for the generators $A_k$, $G_l$ in Onsager's algebra \cite{Davies}.}
\beqa
&&-\frac{(q^{1/2}-q^{-1/2})}{k_+k_-}\omega_0^{(N)}W_{0}^{(N)}+\sum^{N}_{k=1}C_{-k+1}^{(N)}W_{-k}^{(N)} + \epsilon^{(N)}_{+}I\!\!I^{(N)}=0\ ,\nonumber\\
&&-\frac{(q^{1/2}-q^{-1/2})}{k_+k_-}\omega_0^{(N)}W_{1}^{(N)}+\sum^{N}_{k=1}C_{-k+1}^{(N)}W_{k+1}^{(N)} + \epsilon^{(N)}_{-}I\!\!I^{(N)}=0\ .\label{c2}
\eeqa
Also, linear relations for the elements ${\textsf G}_{k+1},{\tilde{\textsf G}}_{k+1}$ can be derived using (\ref{rep}) in (\ref{c2}). They read
\beqa
&&-\frac{(q^{1/2}-q^{-1/2})}{k_+k_-}\omega_0^{(N)}G_{1}^{(N)}+\sum^{N}_{k=1}C_{-k+1}^{(N)}G_{k+1}^{(N)}=0\ ,\nonumber\\
&&-\frac{(q^{1/2}-q^{-1/2})}{k_+k_-}\omega_0^{(N)}{\tilde G}_{1}^{(N)}+\sum^{N}_{k=1}C_{-k+1}^{(N)}{\tilde G}_{k+1}^{(N)}=0\ .\label{c4}\
\eeqa

Having the Sklyanin operator in the form (\ref{SN}) with (\ref{KN}), let us apply above results to the XXZ open spin chain. Plugging (\ref{SN}) with (\ref{KN}) in (\ref{tXXZ}), the Lax operators in (\ref{Kdressed}) which characterize the contribution of the ``bulk'' site$-k$ spin degrees of freedom in (\ref{Kdressed}) reduce to $R-$matrices of the form (\ref{R}) and denoted above $R_{0k}(u)$. As a consequence, the expansion of the transfer matrix (\ref{tXXZ}) at $u=1$ automatically leads to {\it local} conserved quantities \cite{Sklya}. Setting $j_k=1/2$ and $v_k=1$ for all $k=1,...,N$ (homogeneous case), the Sklyanin operator associated with the XXZ open spin chain which enters in (\ref{tXXZ}) leads to consider the nonlocal operators defined by 
\beqa
{\cal W}^{(N)}_{-l}&\equiv&(\otimes_{k=1}^{N}\pi^{(\frac{1}{2})})[{\textsf W}_{-l}^{(N)}]|_{v_k=1}\ ,\qquad
 {\cal W}^{(N)}_{l+1}\equiv(\otimes_{k=1}^{N}\pi^{(\frac{1}{2})})[{\textsf W}_{l+1}^{(N)}]|_{v_k=1}\ ,\nonumber\\
 {\cal G}^{(N)}_{l+1}&\equiv&(\otimes_{k=1}^{N}\pi^{(\frac{1}{2})})[{\textsf G}_{l+1}^{(N)}]|_{v_k=1}\ ,\qquad
\ \  {\tilde{\cal G}}^{(N)}_{l+1}\equiv(\otimes_{k=1}^{N}\pi^{(\frac{1}{2})})[{\tilde{\textsf G}}_{l+1}^{(N)}]|_{v_k=1}\ ,\label{opXXZ}  
\eeqa
for $l\in 0,...,N-1$. Using the two-dimensional representation
\beqa
\pi^{(1/2)}[S_\pm]=\sigma_\pm \qquad \mbox{and}\qquad \pi^{(1/2)}[s_3]=\sigma_3/2\ ,\label{2dim}
\eeqa
their explicit expressions can be easily obtained from (\ref{rep}). For instance, the simplest nonlocal operators take a rather simple form:  
\beqa
{\cal W}^{(N)}_0&=& (k_+\sigma_+ + k_-\sigma_-)\otimes I\!\!I^{(N-1)} + q^{\sigma_3/2}\otimes {\cal W}_0^{(N-1)}\ ,\nonumber \\
{\cal W}^{(N)}_1&=& (k_+\sigma_+ + k_-\sigma_-)\otimes I\!\!I^{(N-1)} + q^{-\sigma_3/2}\otimes {\cal W}_1^{(N-1)}\ ,\label{op}
\eeqa
with ``initial'' conditions 
\beqa
{\cal W}^{(1)}_0&=& k_+\sigma_+ + k_-\sigma_- +  \epsilon_+^{(0)} q^{\sigma_3/2}\ ,\nonumber \\
{\cal W}^{(1)}_1&=& k_+\sigma_+ + k_-\sigma_- +  \epsilon_-^{(0)}q^{-\sigma_3/2}\ .\nonumber
\eeqa

Consequently, the Sklyanin operator (\ref{Kdressed}) associated with the homogenous XXZ open spin chain takes the form (\ref{SN}) with (\ref{KN}), where the non-abelian algebraic structure - originally hidden in the reflection equation (\ref{RE}) - is now encoded in the nonlocal operators (\ref{opXXZ}). For special boundary conditions, note that the nonlocal operators (\ref{op}) coincide with the ones obtained in \cite{Doikou,qDG}.

\subsection{Non-abelian symmetry:  $q-$deformed Onsager's and tridiagonal algebras}
The integrability of the XXZ open spin chain is a consequence of the reflection equation (\ref{RE}) or, more precisely its underlying non-abelian symmetry. Indeed, as shown in \cite{Sklya} transfer matrices of the form (\ref{tXXZ}) are mutually commuting provided $K^{(N)}(u)$ solves (\ref{RE}). As the algebraic content of the Sklyanin operator $K^{(N)}(u)$ is now solely encoded in the finite set of nonlocal operators (\ref{opXXZ}), our next task is to identify the algebra on which they close. As a consequence, this non-abelian algebra ensures the integrability of the XXZ open spin chain.\vspace{1mm} 

To exhibit non-trivial commutation relations among the nonlocal operators (\ref{opXXZ}), one can consider the asymptotic expansion around $u\rightarrow \infty$ of the reflection equation (\ref{RE}) for an arbitrary spectral parameter $v$. In this limit one easily gets from (\ref{RE}) and (\ref{SN}) with (\ref{KN}) the simple intertwinning relations \cite{qDG,qOns}
\beqa
(\pi^{(1/2)}\times id^{(N)})\big[{\textsf W}_0^{(N+1)}\big]|_{v_{N+1}\equiv v^{-1}}K^{(N)}(v) &=&  (\pi^{(1/2)}\times id^{(N)})\big[{\textsf W}_0^{(N+1)}\big]|_{v_{N+1}\equiv v}K^{(N)}(v)\ ,\nonumber\\
(\pi^{(1/2)}\times id^{(N)})\big[{\textsf W}_1^{(N+1)}\big]|_{v_{N+1}\equiv v^{-1}}K^{(N)}(v) &=&  (\pi^{(1/2)}\times id^{(N)})\big[{\textsf W}_1^{(N+1)}\big]|_{v_{N+1}\equiv v}K^{(N)}(v)\ .\label{interfin}
\eeqa
Using the explicit expressions (\ref{rep}) together with (\ref{2dim}), explicitely one finds
\beqa
(\pi^{(1/2)}\times id^{(N)})\big[{\textsf W}_0^{(N+1)}\big]_{v_{N+1}\equiv v} &=&\left(
\begin{array}{cc}
 q^{{1\over 2}}{\textsf W}_0^{(N)}    & k_+v\\
k_-v^{-1}   &  q^{-{1\over 2}}{\textsf W}_0^{(N)}  \\
\end{array} \right) \ ,\nonumber\\
 (\pi^{(1/2)}\times id^{(N)})\big[{\textsf W}_1^{(N+1)}\big]_{v_{N+1}\equiv v} &=&\left(
\begin{array}{cc}
 q^{-{1\over 2}}{\textsf W}_1^{(N)}   & k_+v^{-1}\\
k_-v   &  q^{{1\over 2}}{\textsf W}_1^{(N)}  \\
\end{array} \right) \ .\nonumber
\eeqa
It is then easy to identify the symmetry behind the reflection equation \cite{qDG}.
As an intermediate step, the following algebraic relations\,\footnote{Setting $v=\exp(\mu\lambda)$, $q^{1/2}=\exp(i\mu)$ and choosing the special case $k_+=k_-=1$ one recognizes the relations proposed in \cite{Doikou}.} among the entries of (\ref{KN}) are immediatly obtained from (\ref{interfin}):
\beqa
&&\big[{\textsf W}_0^{(N)},{\cal A}^{(N)}(v)\big] = q^{-1/2}v^{-1}\big(k_-{\cal B}^{(N)}(v) - k_+{\cal C}^{(N)}(v)\big)\ ,\nonumber\\
&&\big[{\textsf W}_0^{(N)},{\cal D}^{(N)}(v)\big] = -q^{1/2}v\big(k_-{\cal B}^{(N)}(v) - k_+{\cal C}^{(N)}(v)\big)\ ,\nonumber\\
&&\big[{\textsf W}_0^{(N)},{\cal B}^{(N)}(v)\big]_q = k_+\big(v{\cal A}^{(N)}(v) - v^{-1}{\cal D}^{(N)}(v)\big)\ ,\nonumber\\
&&\big[{\textsf W}_0^{(N)},{\cal C}^{(N)}(v)\big]_{q^{-1}} = -k_-\big(v{\cal A}^{(N)}(v)-v^{-1}{\cal D}^{(N)}(v) \big)\ ,\label{al}
\eeqa
where the $q-$commutator  \ $[X,Y]_{q}=q^{1/2}XY-q^{-1/2}YX$ \  has been introduced.
Similar relations are obtained for ${\textsf W}_1^{(N)}$, provided one substitutes $q\rightarrow q^{-1},v\rightarrow v^{-1}$ in (\ref{al}). Then, plugging the explicit form (\ref{KN}) of the elements ${\cal A}^{(N)}(v)$, ${\cal B}^{(N)}(v)$, ${\cal C}^{(N)}(v)$, ${\cal D}^{(N)}(v)$ in these relations, $q-$deformed commutation relations for the nonlocal operators (\ref{opXXZ}) are easily obtained. It yields to
\beqa
&&\big[{\cal W}^{(N)}_0,{\cal W}^{(N)}_{k+1}\big]=\big[{\cal W}^{(N)}_{-k},{\cal W}^{(N)}_{1}\big]=\frac{\big({\tilde{\cal G}^{(N)}_{k+1} } - {{\cal G}^{(N)}_{k+1}}\big)}{(q^{1/2}+q^{-1/2})}\ ,\nonumber\\
&&\big[{\cal W}^{(N)}_0,{\cal G}^{(N)}_{k+1}\big]_q=\big[{\tilde{\cal G}^{(N)}}_{k+1},{\cal W}^{(N)}_{0}\big]_q=\rho{\cal W}^{(N)}_{-k-1}-\rho{\cal W}^{(N)}_{k+1}\ ,\nonumber\\
&&\big[{\cal G}^{(N)}_{k+1},{\cal W}^{(N)}_{1}\big]_q=\big[{\cal W}^{(N)}_{1},{\tilde{\cal G}^{(N)}}_{k+1}\big]_q=\rho{\cal W}^{(N)}_{k+2}-\rho{\cal W}^{(N)}_{-k}\ ,\nonumber\\
&&\big[{\cal W}^{(N)}_0,{\cal W}^{(N)}_{-k}\big]=0\ ,\quad 
\big[{\cal W}^{(N)}_1,{\cal W}^{(N)}_{k+1}\big]=0\ \quad \label{qOns}
\eeqa
and
\beqa
&&\big[{\cal G}^{(N)}_{k+1},{\cal G}^{(N)}_{l+1}\big]=0\ ,\quad   \big[{{\tilde{\cal G}}^{(N)}}_{k+1},{\tilde{\cal G}}^{(N)}_{l+1}\big]=0\ ,\quad
\big[{\tilde{\cal G}^{(N)}}_{k+1},{\cal G}^{(N)}_{l+1}\big]
+\big[{\cal G}^{(N)}_{k+1},{\tilde{\cal G}}^{(N)}_{l+1}\big]=0\ \nonumber
\eeqa
for $k,l\in 0,...,N-1$, where
\beqa
\rho = (q^{1/2}+q^{-1/2})^2 k_+k_-\ .\label{rho}
\eeqa\nonumber
Comparing (\ref{qOns}), (\ref{rho}) with \cite{qOns} (eq. $(4)$) we find that the operators ${\cal W}^{(N)}_{-k}$, ${\cal W}^{(N)}_{k+1}$, ${\cal G}^{(N)}_{k+1}$, ${\tilde{\cal G}}^{(N)}_{k+1}$ induce a representation of the $q-$deformed analogue of the Onsager's algebra. We now explain how  ${\cal W}^{(N)}_0,{\cal W}^{(N)}_1$ give a representation of the tridiagonal algebra\,\footnote{It is a coideal subalgebra of $U_{q^{1/2}}(\widehat{sl_2})$ \cite{qDG,TriDiag}. For higher rank quantum Lie algebras, similar relations can also be derived.} recently introduced by Terwilliger \cite{Ter}. By [\cite{Ter}, Definition 3.9] the {\it tridiagonal algebra} ${\mathbb T}$ is the unital associative algebra generated by two symbols $A,B$ subject to the relations
\beqa
\big[A,\big[A,\big[A,B\big]_q\big]_{q^{-1}}\big]&=&\rho \big[A,B\big]\label{tri1}\ ,\\
\big[B,\big[B,\big[B,A\big]_q\big]_{q^{-1}}\big]&=&\rho \big[B,A\big]\label{tri2}\ .
\eeqa
The equations (\ref{tri1}), (\ref{tri2}) are called the {\it tridiagonal} relations or the {\it q-deformed Dolan Grady} relations. Based on \cite{qDG,TriDiag} and previous results one has
\beqa \big[{\cal W}^{(N)}_0,\big[{\cal W}^{(N)}_0,\big[{\cal W}^{(N)}_0,{\cal
W}^{(N)}_1\big]_q\big]_{q^{-1}}\big]&=&\rho\big[{\cal W}^{(N)}_0,{\cal W}^{(N)}_1\big]\ ,\nonumber\\
\big[{\cal W}^{(N)}_1,\big[{\cal W}^{(N)}_1,\big[{\cal W}^{(N)}_1,{\cal
W}^{(N)}_0\big]_q\big]_{q^{-1}}\big]&=&\rho\big[{\cal W}^{(N)}_1,{\cal W}^{(N)}_0\big]\ .
\label{Talg}
\eeqa
Comparing (\ref{tri1})-(\ref{Talg}) we find ${\cal W}^{(N)}_0,{\cal W}^{(N)}_1$ induce a representation of ${\mathbb T}$. We remark that for $N=1,2$ each of ${\cal W}^{(N)}_{-k}$, ${\cal W}^{(N)}_{k+1}$, ${\cal G}^{(N)}_{k+1}$, ${\tilde{\cal G}}^{(N)}_{k+1}$ is a polynomial in ${\cal W}^{(N)}_0,{\cal W}^{(N)}_1$ as shown in \cite{qOns}. From these comments, we see a close relationship between ${\mathbb T}$ and the algebra (\ref{qOns}). The exact nature of this relationship is unclear and needs further analysis. Indeed, it might be possible to construct a different basis of nonlocal operators, say ${\cal A}_k$, ${\cal G}_l$, ${\tilde{\cal G}}_l$, which $q-$commutator relations would be exactly  in one-to-one correspondence with the defining relations of the Onsager's algebra. Although finding such basis is an interesting problem, we leave it for future work.\vspace{1mm}
 
Finally, let us mention that in the context of quantum integrable systems with boundaries the algebraic structure (\ref{Talg}) discovered in \cite{qDG,TriDiag} provides the first {\it explicit} example of boundary quantum algebra.

\subsection{The symmetry of the XXZ open spin chain}
Due to the simplified expression (\ref{SN}) with (\ref{KN}) for the Sklyanin operator it is now possible to expand the transfer matrix in terms  of linear combinations of the nonlocal operators (\ref{opXXZ}). After some straightforward calculations, the transfer matrix (\ref{tXXZ}) becomes
\beqa
t_{XXZ}(u)= \sum_{k=0}^{N-1}{\cal F}_{2k+1}(u)\ {{\cal I}}_{2k+1}^{(N)} + {\cal F}_0(u)\ I\!\!I^{(N)}\label{tfin}
\eeqa
with
\beqa
{{\cal I}}_{2k+1}^{(N)}=\kappa {\cal W}^{(N)}_{-k} + \kappa^* {\cal W}^{(N)}_{k+1} + \frac{\kappa_+}{k_+} {\cal G}^{(N)}_{k+1} 
+ \frac{\kappa_-}{k_-} {\tilde{\cal G}}^{(N)}_{k+1}\ .\label{Ifin}
\eeqa
In (\ref{tfin}), the Laurent polynomials are given by
\beqa
{\cal F}_{2k+1}(u)&=&\frac{(qu^4+q^{-1}u^{-4}-q-q^{-1})}{(q+q^{-1}-u^2-u^{-2})^N}{\overline P}_{-k}^{(N)}(u)\ ,\nonumber\\
{\cal F}_{0}(u)&=&\frac{(qu^4+q^{-1}u^{-4}-q-q^{-1})(q^{1/2}+q^{-1/2})}{(q+q^{-1}-u^2-u^{-2})^N}\Big(\frac{\kappa_+}{k_+}+\frac{\kappa_-}{k_-}\big)\Big(\frac{k_+k_-(q^{1/2}u^2+q^{-1/2}u^{-2})}{(q^{1/2}-q^{-1/2})}{\overline P}_0^{(N)}(u)+ {\overline\omega}_0^{(N)}\Big)\nonumber\\
&&\ \ + \ \frac{\big((q^{1/2}u^2+q^{-1/2}u^{-2})\kappa+ (q^{1/2}+q^{-1/2})\kappa^*\big)\epsilon_+^{(N)}}{(q+q^{-1}-u^2-u^{-2})^N}\ \nonumber\\
&&\ \ + \ \frac{\big((q^{1/2}u^2+q^{-1/2}u^{-2})\kappa^*+ (q^{1/2}+q^{-1/2})\kappa\big)\epsilon_-^{(N)}}{(q+q^{-1}-u^2-u^{-2})^N}\ \nonumber
\eeqa
with the identifications\,\footnote{For the homogenous case and spin$-\frac{1}{2}$ representation at all sites, one simplifies
\beqa
\sum_{l_1<...<l_{N-k}=1}^{N}\alpha_{l_1}\cdot \cdot \cdot\alpha_{l_{N-k}}=
\Big(\frac{2(q+q^{-1})}{(q^{1/2}+q^{-1/2})}\Big)^{N-k-1}\frac{(N-1)!}{(k)!(N-k-1)!} \Big(\frac{2(q+q^{-1})}{(q^{1/2}+q^{-1/2})} \frac{N}{N-k} + \frac{\epsilon^{(0)}_{+}\epsilon^{(0)}_{-}(q^{1/2}-q^{-1/2})^2}{k_+k_-(q^{1/2}+q^{-1/2})}\Big)\ .
\ \nonumber
\eeqa}
\beqa
{\overline P}_{-k}^{(N)}(u)\equiv P_{-k}^{(N)}(u)|_{\{j_l=\frac{1}{2};v_l=1\ \forall l\}} \qquad \mbox{and}\qquad {\overline \omega}_0^{(N)}\equiv {\omega}_0^{(N)}|_{\{j_l=\frac{1}{2};v_l=1\ \forall l\}}\ .\nonumber
\eeqa

As mentionned above, the reflection equation (\ref{RE}) implies the commutativity of the transfer matrices of the form (\ref{tXXZ}) which is nothing but a manifestation of the abelian symmetry of the XXZ open spin chain. Indeed, using the properties of the reflection equation and the Yang-Baxter algebra for arbitrary spectral parameters $u,v$ it is not difficult to show that \cite{Sklya}
\beqa
\big[t_{XXZ}(u),t_{XXZ}(v)\big]=0\ .\label{inv}
\eeqa
As an immediate consequence, the nonlocal operators ${{\cal I}}_{2k+1}^{(N)}$, $k=0,...,N-1$ as defined in (\ref{Ifin}) generate an abelian subalgebra of the $q-$deformed analogue of the Onsager's algebra: replacing the expansion (\ref{tfin}) in (\ref{inv}) we deduce that all nonlocal operators (\ref{Ifin}) are in involution i.e.
\beqa
\big[{\cal I}^{(N)}_{2k+1},{\cal I}^{(N)}_{2l+1}\big]=0 \qquad \mbox{for all} \qquad k,l\in 0,...,N-1\ \label{sub}
\eeqa
for generic values of the left and right boundary parameters. As a consistency check, note that it is possible to derive the same result using solely the defining relations (\ref{qOns}). Also, for $q=1$ and $\kappa_\pm=0$
 the hierarchy of nonlocal operators ${{\cal I}}_{2k+1}^{(N)}$, $k=0,...,N-1$ coincides with the hierarchy discovered by Onsager in the Ising model \cite{Ons} or, equivalently \cite{Perk,Davies}, the one discovered by Dolan and Grady in \cite{DG}.\vspace{1mm}

Finally, combining the expansion of the transfer matrix in terms of {\it local} conserved quantities together with the expansion in terms of {\it nonlocal} ones, we deduce the abelian symmetries of the model.
Using  (\ref{expH}) in (\ref{inv}) together with the expansion (\ref{tfin}) for $u\rightarrow v$ we end up with
\beqa
\big[H_{XXZ},{\cal I}^{(N)}_{2k+1}\big]=0 \qquad \mbox{for all} \qquad k\in 0,...,N-1\ .\label{abel}
\eeqa
Obviously, it is possible to restrict the boundary parameters to special values. 
For instance, for $\kappa^*=0$ and $\kappa_\pm=0$, the Hamiltonian commutes with all nonlocal operators  ${\cal W}^{(N)}_{-k}$, $k=0,...,N-1$. In particular, the case $k=0$ was identified in \cite{Doikou}.

\section{Comments}
We have shown that the integrability of the XXZ open spin chain with general integrable boundary conditions relies on the existence of nonlocal operators of the form (\ref{opXXZ}) generating a $q-$deformed analogue of the Onsager's Lie algebra with defining relations (\ref{qOns}). The transfer matrix of the model has been expanded in terms of a hierarchy of mutually commuting quantities (\ref{Ifin}) which generate the abelian symmetry (\ref{abel}) of the model. In particular, these quantities - which generalize the ones of Onsager \cite{Ons} or Dolan-Grady \cite{DG} - are expressed as linear combinations of the nonlocal operators (\ref{opXXZ}) with coefficients depending on the boundary parameters.\vspace{1mm}

Clearly, the results presented here open the possibility to study the XXZ open spin chain - and more generally quantum integrable models with boundaries - from a new point of view, based on Onsager's approach to integrable lattice systems.
To motivate the proposal, let us first recall the ``Onsager's approach''. 
In the pioneering work of Onsager \cite{Ons} for the two-dimensional classical Ising model in zero magnetic field, the introduction of the so-called Onsager's Lie algebra with generators $A_k,G_l$ and defining relations
\beqa
[A_k,A_l]=4G_{k-l}\ , \quad [G_l,A_k]= 2A_{l+k}-2A_{-l+k}\ , \quad [G_k,G_l]=0\ \label{Ons}
\eeqa
for any integers $k,l$ played a crucial role in the analysis. Actually, Onsager was able to show that the transfer matrix of the Ising model can be written in terms of the fundamental generators of (\ref{Ons}). For an arbitrary parameter $\lambda$, it takes the simple form\,\footnote{In original Onsager's paper \cite{Ons} the parameter $\lambda=1/2$ or $\lambda=1$.} (see \cite{Davies})
\beqa
t_{Ising}(\lambda)=\exp(\lambda a_0A_0)\exp(\lambda a_1A_1)\exp((1-\lambda)a_0A_0)\ \label{T}\ ,
\eeqa
where $a_0,a_1$ are parameters depending on the temperature of the system and the coupling constant in each direction of the two-dimensional lattice. For a rectangular lattice of size $L\times N$ (with site index ``$i$'' in one direction), the operators can be represented by matrices of dimension $2^N$ which read \cite{Ons,Davies}:
\beqa
A_0=\sum_{j=1}^{N-1}\sigma^j_{1}\sigma^{j+1}_{1} \pm \sigma^N_{1}\sigma^{1}_{1} \qquad \mbox{and}\qquad   A_1=\sum_{j=1}^{N}\sigma^j_{3}\label{A0A1Ising}\ .
\eeqa
Here the choice of sign in the definition of $A_0$ corresponds to periodic/antiperiodic ($+/-$) boundary conditions in one direction of the lattice. For any boundary conditions, it is known that $A_0,A_1$ satisfy the Dolan-Grady relations \cite{DG}
\beqa
[A_0,[A_0,[A_0,A_1]]]=\rho_0[A_0,A_1]\qquad \mbox{and}\qquad [A_1,[A_1,[A_1,A_0]]]=\rho_0[A_1,A_0]\ ,\label{relDG}
\eeqa
where $\rho_0=16$. Although these relations clearly follow from (\ref{Ons}), it is important to mention that these relations are actually {\it sufficient} to generate the Onsager's algebra \cite{Perk,Davies,Ro} and imply the integrability of the Ising model with transfer matrix (\ref{T}). 
For generic values of $a_0,a_1$, using the Dolan-Grady relations (\ref{relDG}) in the expansion of (\ref{T}) it is possible to show \cite{Davies} that the quantum Hamiltonian
\beqa
H = \coth(2a_1)A_0 + \coth(2a_0)A_1 + \frac{\sinh(2-4\lambda)a_0}{\sinh(2a_0)}G_1 - \frac{\sinh(2\lambda a_0)\sinh(2(1-\lambda)a_0)}{\sinh(2a_0)}(A_1-A_{-1})\ \label{Hlambda}
\eeqa
satisfies $[t_{Ising},H]=0$. More generally, any local conserved quantities can be written as linear combination of the fundamental operators $A_k,G_l$ which obey (\ref{Ons}).  Based on the properties of the non-abelian Onsager's algebra (\ref{Ons}), Onsager found the eigenvalues of the transfer matrix in the principal direction which lead to the determination of the free energy. Later on, this approach was applied to a few other examples of quantum integrable spin chains which enjoy the Onsager's symmetry algebra (\ref{Ons}) - Among them, one can cite for instance the XY model \cite{DG}, the superintegrable chiral Potts model \cite{Potts,Davies,Ro,Ro05} and generalizations \cite{Ahn}. For each of these spin chains, the Hamiltonian and more generally any higher local conserved quantity  are simply expressed in terms of operators which form the abelian subalgebra of Onsager's algebra. Remarkably, this finite set of linearly independent operators coincides with (\ref{Ifin}) provided one identifies $q=1$ in (\ref{qOns}) and fix the boundary parameters $\kappa,\kappa^*,\kappa_\pm$ to suitable values.  For all these models, the spectral problem for the Hamiltonian or higher conserved quantities reduces to the one associated with the family (\ref{Ifin}) at $q=1$, which solution is well-known \cite{Ons,Davies,Potts,Ro}.\vspace{1mm}

By analogy, it is thus natural from previous analysis to consider the XXZ open spin$-\frac{1}{2}$ chain with general integrable boundary conditions from that point of view. The transfer matrix of this model (\ref{tXXZ}) being simply expressed in the form (\ref{tfin}) i.e. in terms of the generators (\ref{Ifin}) of the $q-$deformed Onsager's abelian subalgebra (\ref{sub}), any local conserved quantity can be written in terms of (\ref{Ifin}) differentiating \ $t_{XXZ}(u\equiv\exp(\lambda))$ given by (\ref{tfin}) with respect to  $\lambda$ at $\lambda=0$. For instance, using (\ref{expH}) one immediately gets
\beqa
H_{XXZ}&=& \frac{(q^{1/2}-q^{-1/2})}{2(q^{1/2}+q^{-1/2})(\kappa+\kappa^*)(\epsilon_+^{(0)}+\epsilon_-^{(0)})}\Big(\sum_{k=0}^{N-1}\frac{d}{d\lambda}{\cal F}_{2k+1}(u\equiv e^{\lambda})|_{\lambda=0}\ {{\cal I}}_{2k+1}^{(N)} + \frac{d}{d\lambda}{\cal F}_0(u\equiv e^{\lambda})|_{\lambda=0}\ I\!\!I^{(N)}\Big)\ \nonumber \\
&&- \Big(N\Delta + \frac{(q^{1/2}-q^{-1/2})^2}{2(q^{1/2}+q^{-1/2})}\Big)I\!\!I^{(N)}\ .  
\eeqa
For small values of $N$ ($N=1$ \cite{qDG,TriDiag}, $N=2,3$), it is not difficult to solve numerically the spectral problem for ${\cal I}^{(N)}_{2k+1}$ for $k=0,...,N-1$ which exhibits some interesting features depending on
the boundary parameters. For instance, for non-vanishing values of the boundary parameters the spectrum of the simplest quantity ${\cal I}^{(N)}_1$ is non-degenerate. However, degeneracies appear for $\kappa^*=\kappa_\pm=0$ (or equivalently $\kappa=\kappa_\pm=0$). For higher values of $N$ important simplifications in the analysis occur for special relations \cite{Nepo,Cao} between left and right boundary parameters. In this case, the eigenstates of (\ref{Ifin}) can be factorized in agreement with the results of \cite{Cao}. Details will be considered elsewhere.

\vspace{0.3cm}

\noindent{\bf Acknowledgements:} P.B. thanks  V. Rittenberg for discussions, comments on the manuscript, suggestions and interest in this work. The authors thank the referees for comments and suggestions. K. Koizumi is supported by CNRS and French Ministry of Education and Research. Part of this work is supported by the TMR Network EUCLID ``Integrable models and applications: from strings to condensed matter'', contract
number HPRN-CT-2002-00325.\vspace{0.5cm}

\end{document}